\documentclass[pre,twocolumn,superscriptaddress,showpacs]{revtex4}
\usepackage{graphicx,epsfig}
\begin{document}
\title{Time-independent approximations for periodically driven systems with friction}

\author{Saar Rahav}
\affiliation{Department of Physics, Technion, Haifa 32000, Israel.}
\author{Eli Geva}
\affiliation{Department of Physics, Technion, Haifa 32000, Israel.}
\author{Shmuel Fishman}
\affiliation{Department of Physics, Technion, Haifa 32000, Israel.}
\date{29 November 2004}

\begin{abstract}
The classical dynamics of a particle that is driven by a rapidly
oscillating potential (with frequency $\omega$) is studied.
The motion is separated into a slow part and a fast part
that oscillates around the slow part.
The motion of the slow part is found to be described by a time-independent 
equation that is derived as an expansion in orders
of $\omega^{-1}$ (in this paper terms to the order $\omega^{-3}$
are calculated explicitly). This time-independent equation is used
to calculate the attracting fixed points and their basins of
attraction. The results are found to be in excellent agreement
with numerical solutions of the original time-dependent problem.
\end{abstract}

\pacs{45.05.+x,45.50.-j}

\maketitle

 Time-dependent systems typically exhibit behavior which is more complicated 
than the one of the corresponding time-independent ones. 
Moreover, physicists did not develop yet an intuition on the dynamics
of time-dependent systems to the level of the one that exists
for time-independent ones.
As a result, it is 
difficult to predict the qualitative properties of driven systems
even in cases in which it is easy to understand the dynamics of
time-independent systems that are similar.
In this work, the dynamics of some time-dependent 
driven systems will be related to the dynamics of
 time independent ones.
In particular, this will be done for some systems where there is a clear
separation of time scales. This will enable
qualitative and quantitative analysis of the dynamics of
some class of time-dependent systems using the
experience with time-independent ones.

There is a vast literature on methods which approximate a 
dynamical system by a ``simpler'' system with smaller 
number of degrees of freedom. These include, among others,
averaging methods~\cite{lichtenberg,averaging},
multiple time scale analysis~\cite{averaging},
and centre manifold theory~\cite{carr}.
In fact, the method employed in this article can be viewed
as a particular example for the more general method of
multiple time scales analysis~\cite{averaging}.
It should be noted that the method used in this work is tailored
for equations of the form (\ref{newton}), and therefore its
application is much simpler then the use one of the more general 
methods. However, it is representative of many physical
problems.

The more known case of separation of time scales is the adiabatic one.
In this case, the system evolves on a time scale which is much shorter then
the time scale of the driving. There are many works which treat different
adiabatic approximations~\cite{lichtenberg,averaging}. On a qualitative level, one can
understand the resulting dynamics by treating the driving as if it was fixed 
in time, letting the system evolve, and then change the values of the 
parameters of the driving according to their time dependence.

Less known, but not less interesting, is the opposite limit. In this
case, the typical time scales of the dynamics of the system in absence
of the driving are much longer than the period of the driving. 
A remarkable effect of such driving is ``dynamical stabilization'',
in which a particle, that in absence of the driving can escape 
from some region,
may be trapped by the rapidly oscillating field. 
Examples for this phenomenon include
 the Kapitza pendulum~\cite{kapitza} and the Paul trap~\cite{paul}.

A simple treatment of rapidly driven systems is given in textbooks~\cite{LL1},
following Kapitza's work on the inverted pendulum~\cite{kapitza}.
In this approximate calculation the motion is separated into a sum
of a slow part and rapid oscillations around it. The rapid oscillations 
are computed explicitly and their effect on the slow motion is found.
The treatment of Landau and Lifshitz turns out to be the leading
order of an expansion in $\omega^{-1}$
(where $\omega$ is the driving frequency).
It was extended to the order $\omega^{-4}$ 
in recent works~\cite{rahav03a,rahav03b}.
In these works it was demonstrated that for rapidly driven Hamiltonian systems
it is possible to obtain a time independent Hamiltonian that 
controls the slow motion. It is obtained by a canonical transformation
as an expansion in powers of $\omega^{-1}$. 

In this article, the method developed in~\cite{rahav03a,rahav03b} is
extended to rapidly driven classical systems in the presence of
friction where the Hamiltonian formalism is inapplicable,
since energy is dissipated.
The (modest) goal of this article is to demonstrate how
one can understand and predict the qualitative properties
of the dynamics using very simple approximations.
In particular, it will be demonstrated how this method can be
used in order to predict qualitatively the general form
of the basins of attraction for such systems and how to compute
their boundaries from the equation of motion
of the slow part.

Newton's equation of motion, for a rapidly driven system, is given
by
\begin{equation}
\label{newton}
m \ddot{x} + \alpha \dot{x} = -V_0^\prime (x) - V_1^\prime (x, \omega t),
\end{equation}
where $m$ is the mass of the particle and $\alpha$ is the friction
constant. The potential is $V(x, \omega t) = V_0 (x) + V_1 (x, \omega t)$
that is
chosen so that the time average of $V_1$ over a period vanishes.
Derivatives with respect to the coordinate and time of $f(x,t)$
are denoted by $f^\prime$ and $\dot{f}$ respectively,
while $\overline{f}$ denotes the average over a period.

The slow and fast motion are separated with the help of the ansatz
\begin{equation}
\label{ansatz}
x(t)= X(t)+ \xi \left( X, \dot{X}, \tau \right),
\end{equation}
where $\xi$ is periodic in the fast time variable
 $\tau \equiv \omega t$, with a vanishing average
over a period, and
$X(t)$ is the slow part of the motion. 
The fast time $\tau$ is treated as an independent variable.
One ensures that
$X(t)$ is indeed slow by choosing $\xi$ in such a way so that
the equation of motion for $X$ does not depend explicitly on time.
This can be done, at least approximately, by expanding $\xi$ in a power series
in $\omega^{-1}$, using
\begin{equation}
 \xi = \sum_{n=1}^\infty \frac{1}{\omega^n}\xi_n.
\end{equation}
Then, the functions $\xi_n$ are determined, order by order (in $\omega^{-1}$),
from the condition that the remaining equation of motion for $X$
is time-independent. This procedure leads to new
time-independent terms in the equation of motion for $X(t)$.
(These terms cannot be canceled by a choice of $\xi_n$ which stay
bounded for large times.)
An explicit derivation of the first few terms for a
similar problem can be found in~\cite{rahav03b}.
For an equation of motion of the form (\ref{newton}),
the perturbation theory results in
 the following explicit solution for $\xi$ (given in terms of $X(t)$)
\begin{eqnarray}
\label{xi}
\xi \left( X, \dot{X}, \omega t \right) & \simeq & - \frac{1}{m \omega^2} \int^{(2)\tau}\left[ V_1^\prime \right]+ \frac{2 \dot{X}}{m \omega^3} \int^{(3) \tau} \left[ V_1^{\prime \prime} \right] \nonumber \\ & + & \frac{\alpha}{m^2 \omega^3} \int^{(3) \tau}\left[ V_1^\prime \right] + O(\omega^{-4}).
\end{eqnarray} 
Substitution of (\ref{xi}) in (\ref{newton}) leads
to the slow equation of motion, for $X(t)$,
\begin{eqnarray}
\label{slow}
m \ddot{X} & = & - \alpha \dot{X} - V_0^\prime (X) - \frac{1}{m \omega^2} 
\overline{\int^\tau \left[V_1^{\prime \prime} \right]\int^\tau \left[V_1^{\prime} \right]} \nonumber \\ & + & \frac{\alpha}{m \omega^3} \overline{\int^\tau \left[V_1^{\prime \prime} \right]\int^{(2)\tau} \left[V_1^{\prime} \right]}
+O(\omega^{-4}).
\end{eqnarray}
 The symbol $\int^\tau \left[...\right]$ denotes an integral over $\tau$,
defined {\em only} for periodic functions  of $\tau$ with vanishing 
average, which is performed in such a way so that the resulting function
of $\tau$ is also periodic with vanishing average. This integral is easy
to compute using the Fourier expansion of the integrand (see 
\cite{rahav03b} for details). 
Multiple application of the integral ($j$ times) is denoted by $\int^{(j) \tau} \left[...\right]$.
The overline denotes an average over a period
of $\tau$. Therefore, equation (\ref{slow}) does not depend explicitly 
on time (and its solution will {\em not} exhibit oscillations with the external 
frequency $\omega$).

The leading-order ($\omega^{-2}$) correction to the motion in absence of the driving
can be seen as resulting from the
effective potential
\begin{equation}
\label{veff}
V_{eff} (X)= V_0 (X) + \frac{1}{2 m \omega^2} \overline{\left( \int^\tau \left[ V_1^\prime (X, \tau)\right]\right)^2}
\end{equation}
that acts on the slow motion. However, at higher orders, terms appear which do not 
seem to result from a potential.
Note that in spite of the rapid oscillations, the friction,
and hence the energy dissipation, is associated only
with the slow motion (at the order $\omega^{-2}$).

Equations (\ref{xi}) and (\ref{slow}) are the results of a high
frequency perturbation theory. They result in a mapping of a time-dependent
problem into a time-independent one. 
In any given order in $\omega^{-1}$, this theory reproduces
the result of the mathematical theory of separation of time
scales, but it is much simpler.
One of the goals of this article
is to demonstrate how these equations can be used to understand
the dynamics of such driven systems. It will be demonstrated by a simple
example.

\begin{figure}[htb]
\includegraphics[width=7cm,height=5cm]{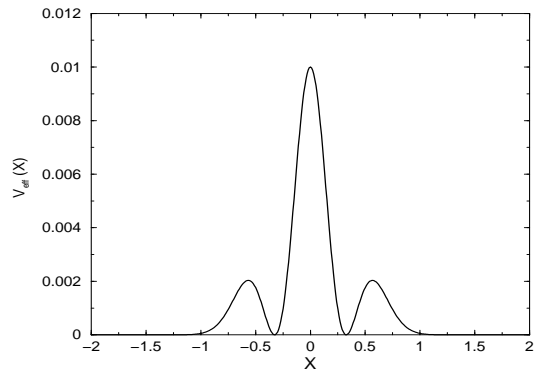}
\caption{The effective potential for the slow motion. The values of the parameters in (\ref{veffexample}) are $A=m=1$, $\beta=4$, $k=2$ and $\omega=10$. \label{fig1}} 
\end{figure}

Consider a particle which, apart from the friction, is under the influence
of an oscillating field given by
\begin{equation}
\label{example}
V_1 (x, \omega t) = A e^{- \beta x^2} \sin (kx) \cos (\omega t + \phi).
\end{equation}
This simple system is of interest since the time average of the potential
vanishes, $V_0(x)=0$. Therefore, the influence of the oscillatory field
is dominant even at high frequencies, in contrast to systems with
$V_0 (x) \ne 0$. According to the high frequency perturbation
theory, the slow part of the motion of the particle can be 
(approximately) viewed
as motion, with friction, in the effective potential
\begin{equation}
\label{veffexample}
V_{eff} (X) = \frac{A^2}{4 m \omega^2} e^{-2 \beta X^2} \left[ k \cos (k X) - 2 \beta X \sin (k X) \right]^2.
\end{equation}
This effective potential is depicted in Fig.~\ref{fig1}.
It exhibits several minima. These are of
interest since a particle moving in a time-independent potential,
in the presence of friction, will be found at one of these minima
after a sufficiently long time, in spite of the fact that
the linear stability of these points is time-dependent in the 
original system.  
While the potential (\ref{veffexample}),
depicted in Fig.~\ref{fig1}, has several minima, two of those, at $x \simeq \pm 0.3266$, 
are more pronounced. Based on these observations, one can predict that
after sufficiently long time the particle will be found in the
vicinity of one of the minima also for the time-dependent system.

\begin{figure}
 \begin{minipage}[c]{0.2\textwidth}
	\centering \includegraphics[width=4cm]{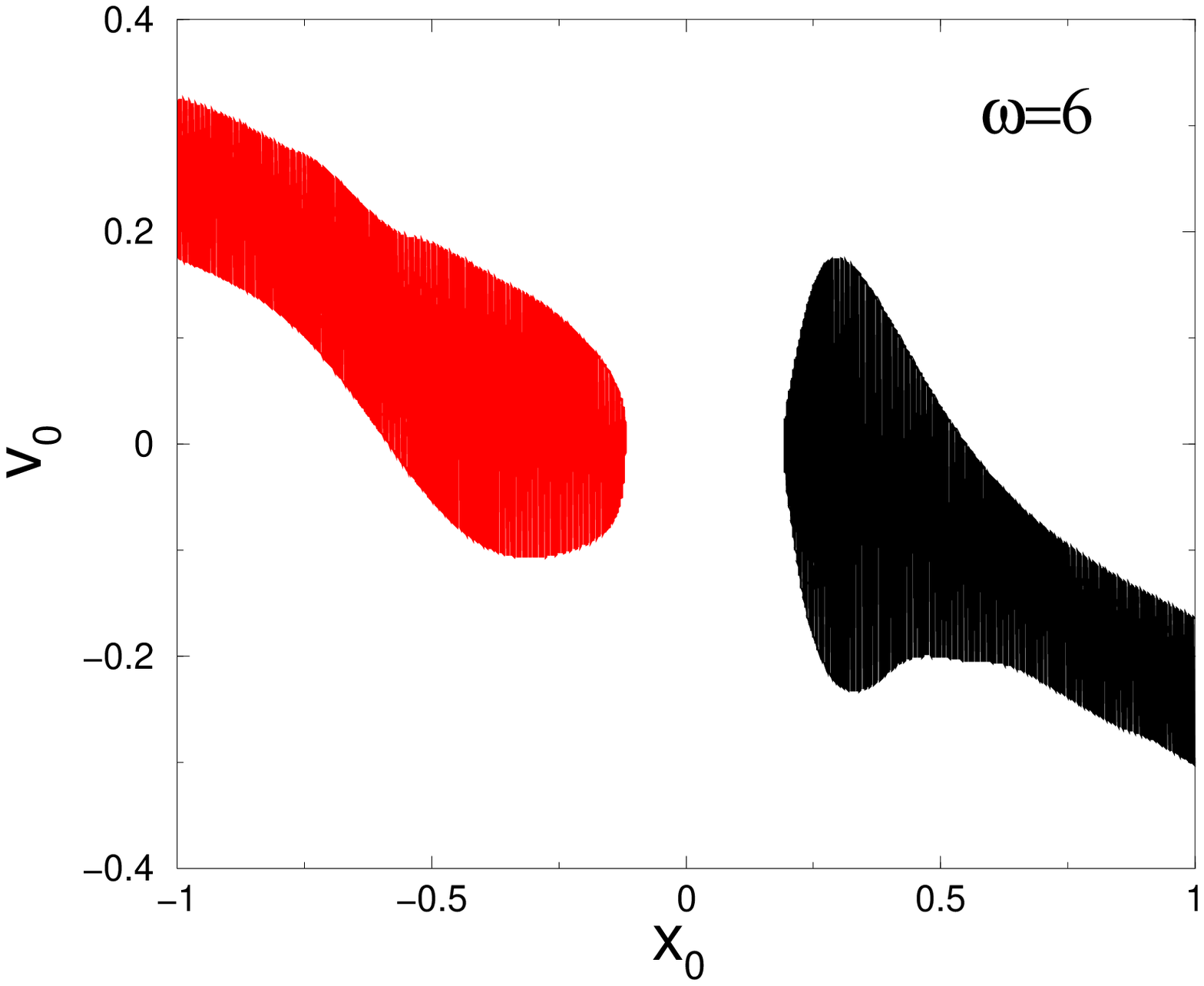}
 \end{minipage}%
 \hspace{0.04\textwidth}%
 \begin{minipage}[c]{0.2\textwidth}
	\centering \includegraphics[width=4cm]{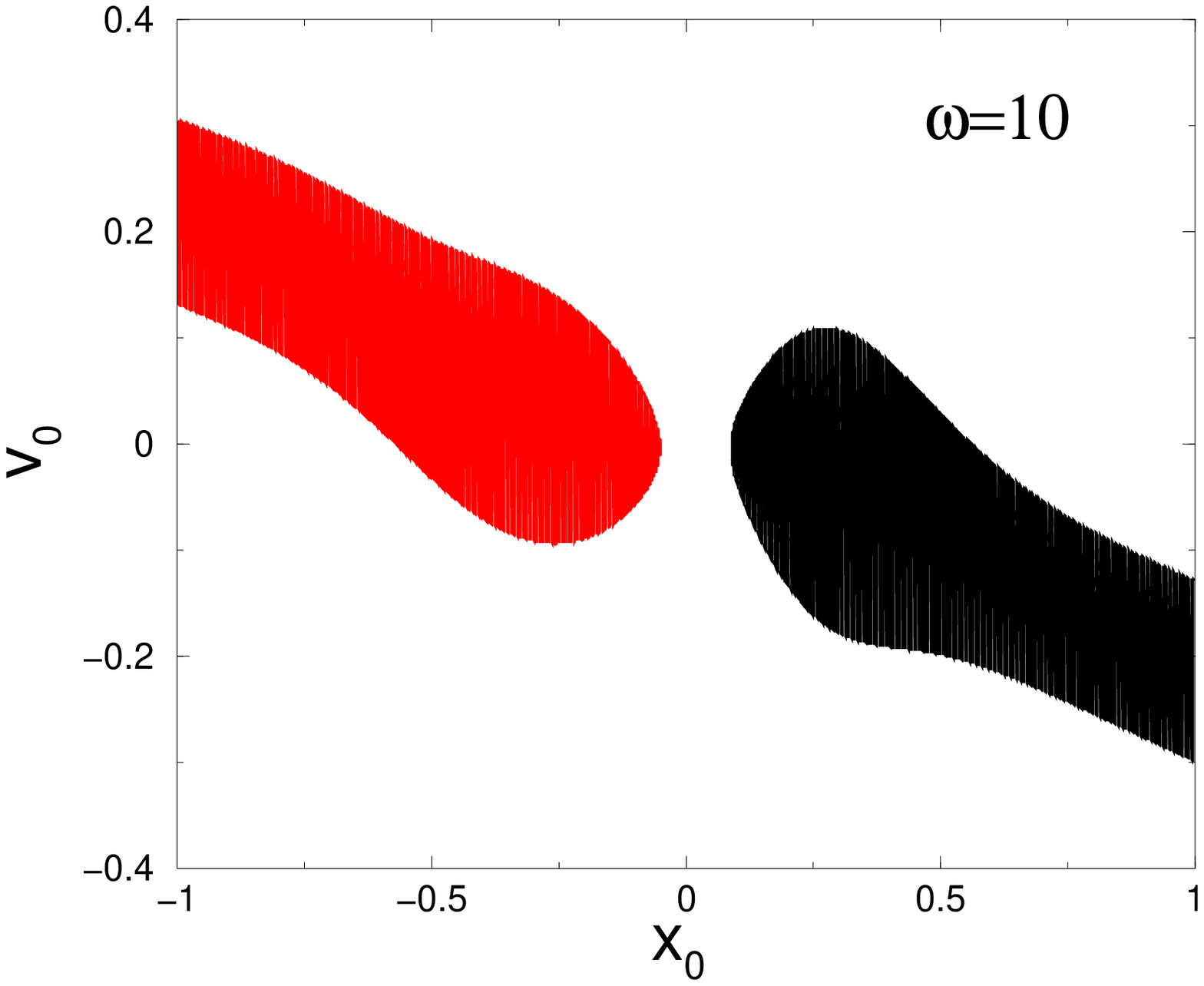}
 \end{minipage}\\[5pt]
 \begin{minipage}[c]{0.2\textwidth}
	\centering \includegraphics[width=4cm]{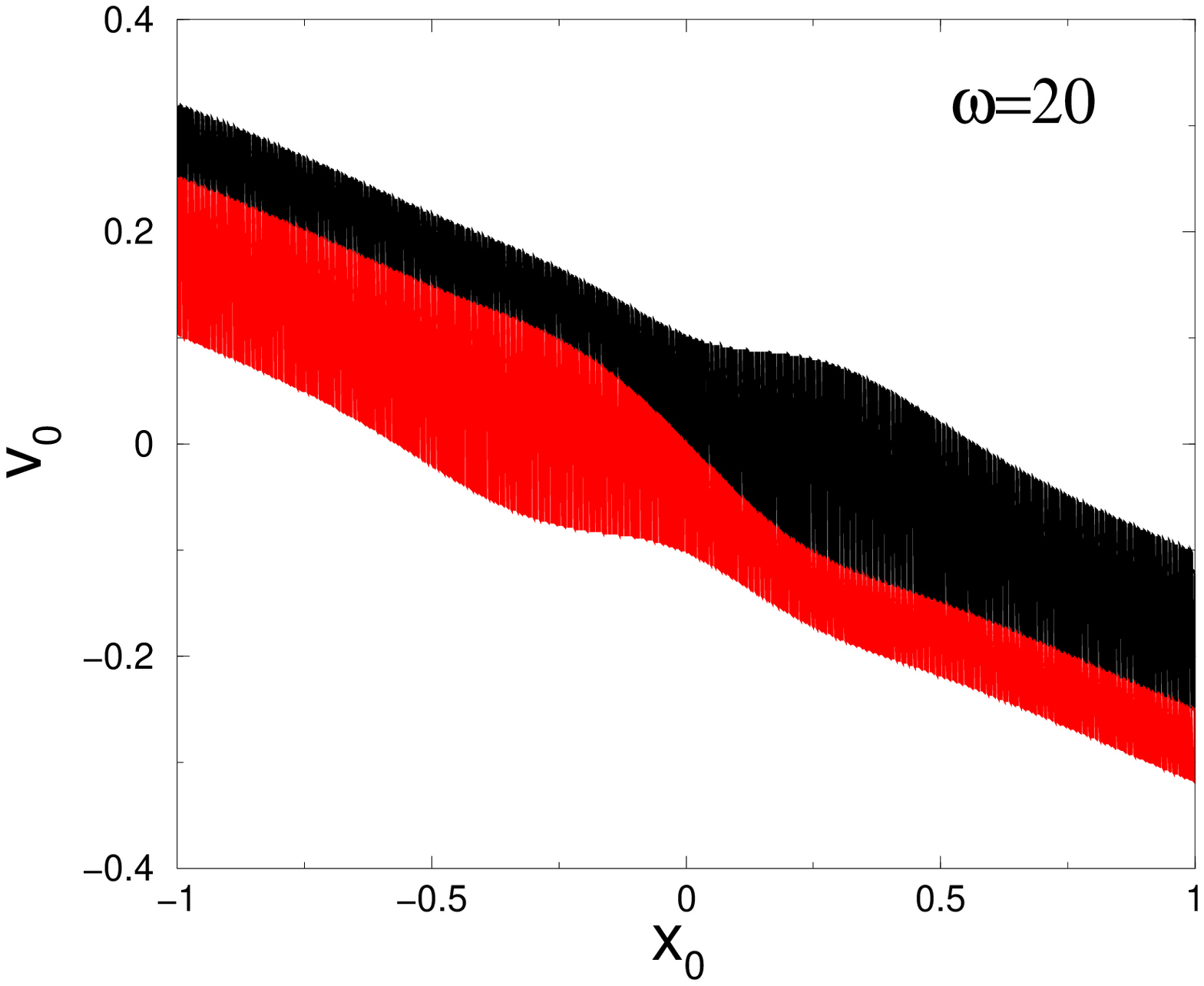}
 \end{minipage}%
 \hspace{0.04\textwidth}%
 \begin{minipage}[c]{0.2\textwidth}
	\centering \includegraphics[width=4cm]{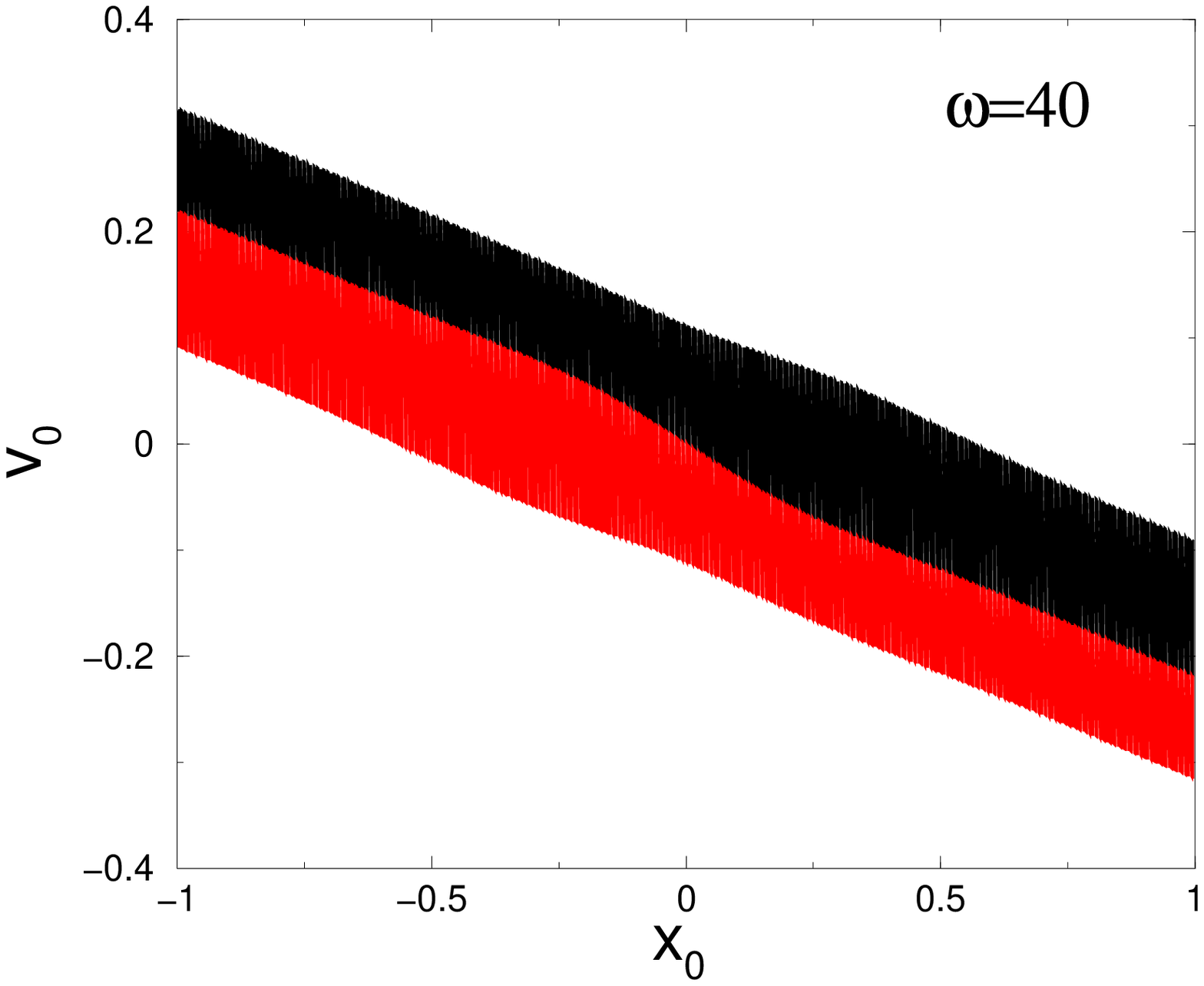}
 \end{minipage}%
 \caption{Basins of attraction of $x(t\rightarrow \infty) \simeq \pm 0.3266$ for the time dependent system (\ref{example}). Results for several frequencies are presented, while $A=m=1$, $\beta=4$, $k=2$ and $\phi=0$.  \label{fig2}}
\end{figure}

In this article, we are interested mainly in the basins of attraction 
of the two minima located at $x \simeq \pm 0.3266$, that is, the initial
values $x_0 \equiv x(0)$,$v_0 \equiv \dot{x}(0)$ which evolve to these minima for long times.
These basins of attraction, for different values of the frequency,
are depicted in Fig.~\ref{fig2}.
The basins of attraction exhibit some qualitative properties which are of 
interest. At lower frequencies ($\omega=6,10$) these basins are separated
while for higher frequencies ($\omega=20,40$) they have a common boundary.
This qualitative behavior can be understood with the help of the effective potential
(Fig.~\ref{fig1}) for the approximate time-independent system. 

Consider the trajectory (of the effective time-independent system) which
starts with the initial values $X_0=0$, $\dot{X}_0=\epsilon>0$
(that is, arbitrarily close to $X_0=\dot{X}_0=0$).
If this trajectory, after a long time, manages to pass over the first
maxima of $V_{eff}$ with positive $X$, then any
other trajectory which starts say with $X_0<0$ and a (large enough) 
positive velocity will also do so. In this case, the basins of attraction
of $x \simeq \pm 0.3266$ are separated. In contrast, if this orbit, 
starting at $X_0=0$ and $\dot{X}_0=\epsilon>0$,
is trapped at $x \simeq 0.3266$ after a long time then the basins do
have a common boundary which includes the point $X_0=\dot{X}_0=0$.
Consider the effective potential depicted in Fig.~\ref{fig1}. A particle 
located near $X=0$ will feel an effective force due to it.
This effective force will accelerate the particle while 
the friction will decelerate it. If the friction is dominant the 
particle will be trapped in the first minimum and it will
be found near $x = 0.3266...$ after a long time. In contrast,
if the effective force is strong enough one can expect that
the particle will be able to pass the first maximum
of $V_{eff} (X)$ at positive $X$ and therefore will not be near $x \simeq 0.3266$ after
 a long time. 
In the example used in the present numerical
investigation the friction does not scale 
with the frequency while the effective potential (and the resulting 
force) scales as $\omega^{-2}$. Therefore, one can expect to find
a transition from separated basins of attraction to basins with a common 
boundary as the frequency is increased.
This is consistent with the numerical results presented in Fig.~\ref{fig2}.

\begin{figure}[h]
 \begin{minipage}[c]{0.2\textwidth}
	\centering \includegraphics[width=4cm]{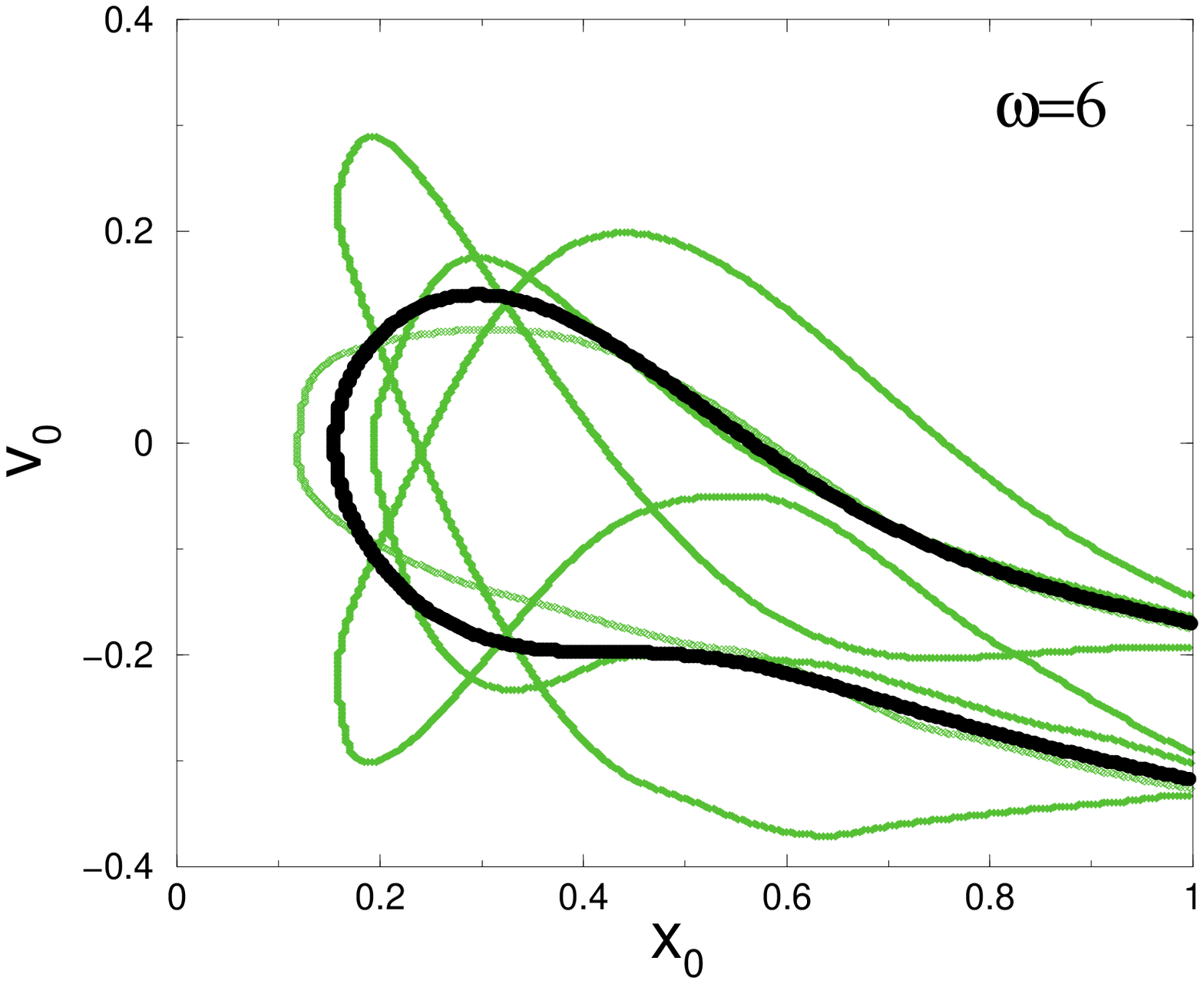}
 \end{minipage}%
 \hspace{0.04\textwidth}%
 \begin{minipage}[c]{0.2\textwidth}
	\centering \includegraphics[width=4cm]{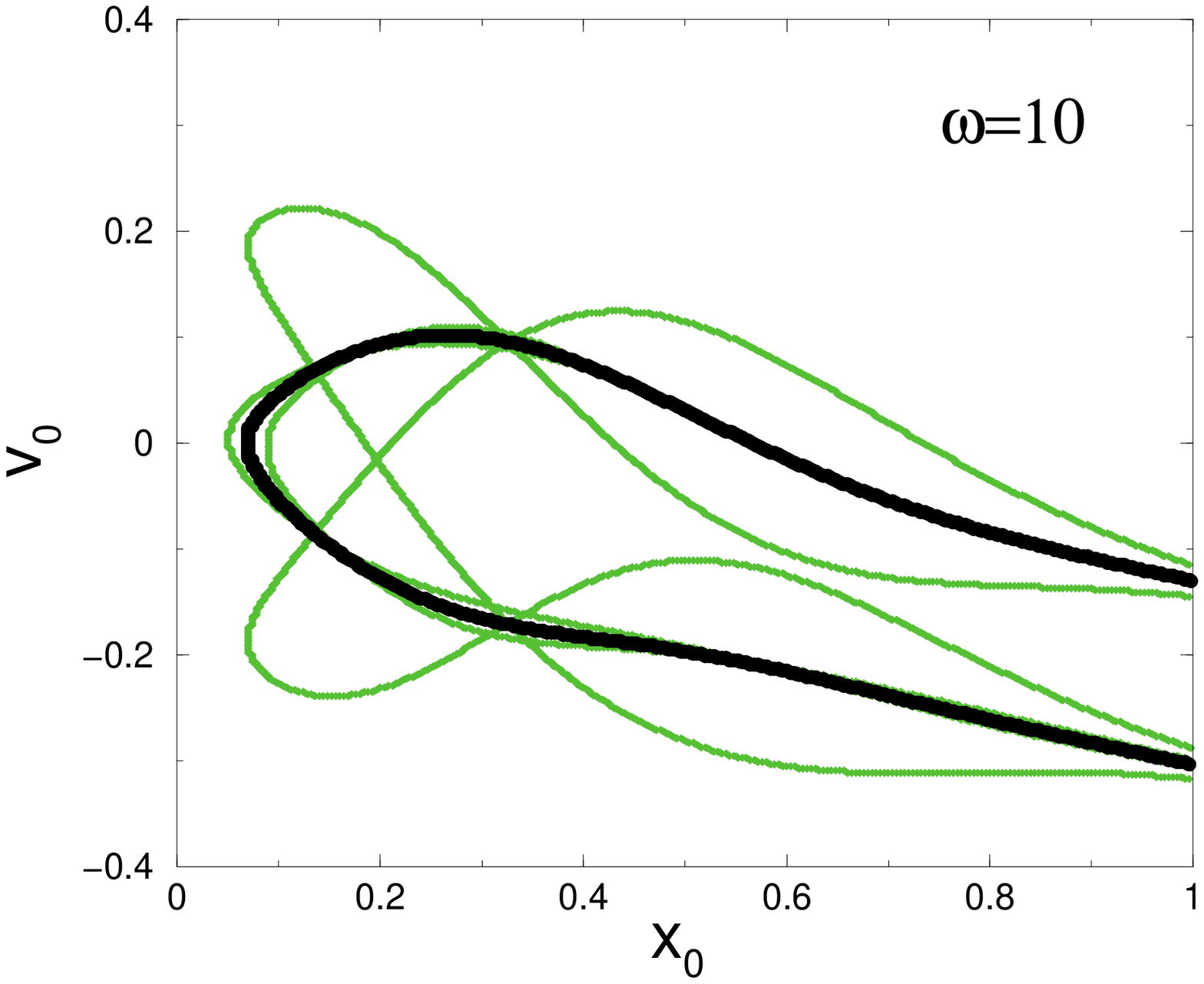}
 \end{minipage}\\[5pt]
 \begin{minipage}[c]{0.2\textwidth}
	\centering \includegraphics[width=4cm]{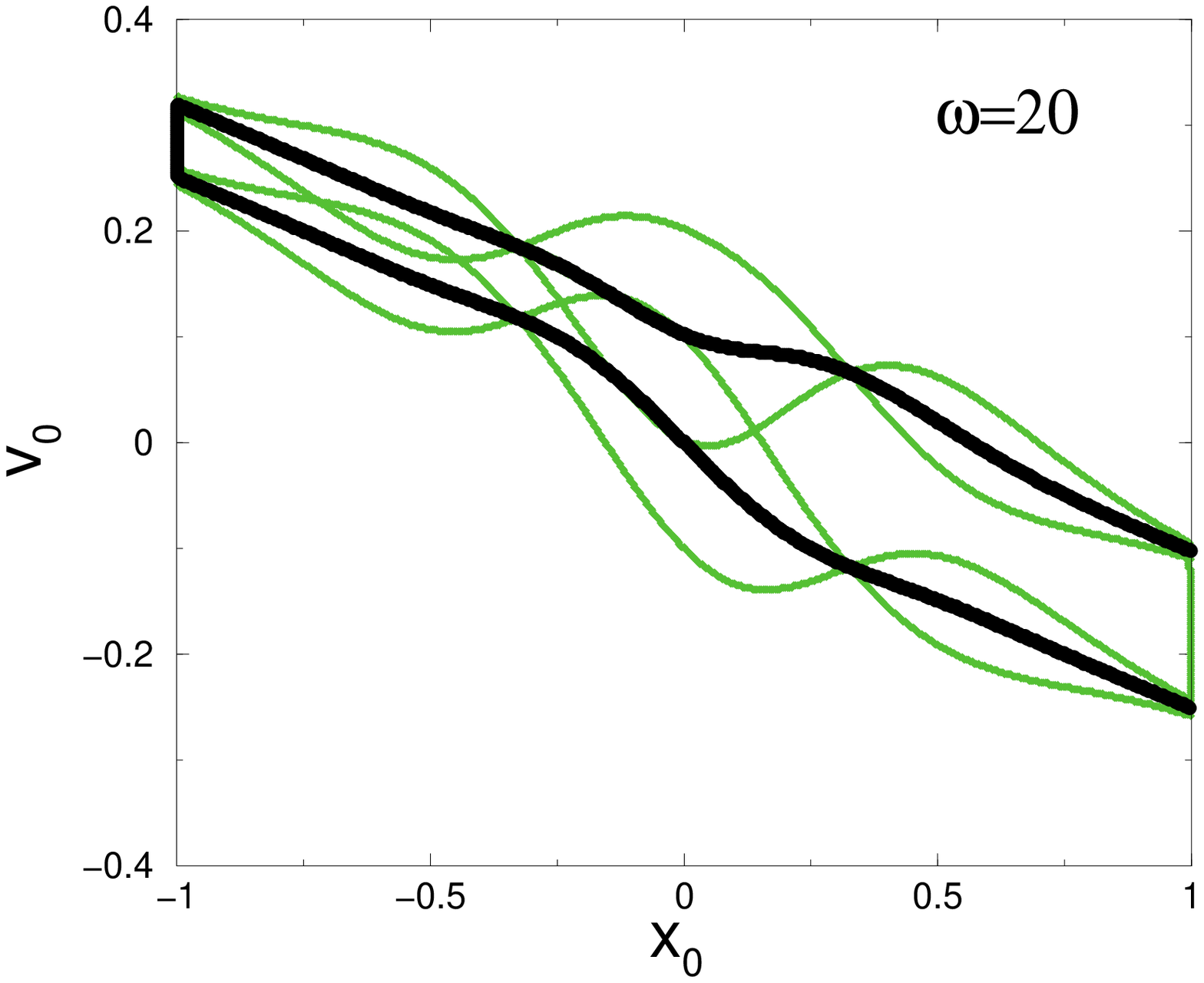}
 \end{minipage}%
 \hspace{0.04\textwidth}%
 \begin{minipage}[c]{0.2\textwidth}
	\centering \includegraphics[width=4cm]{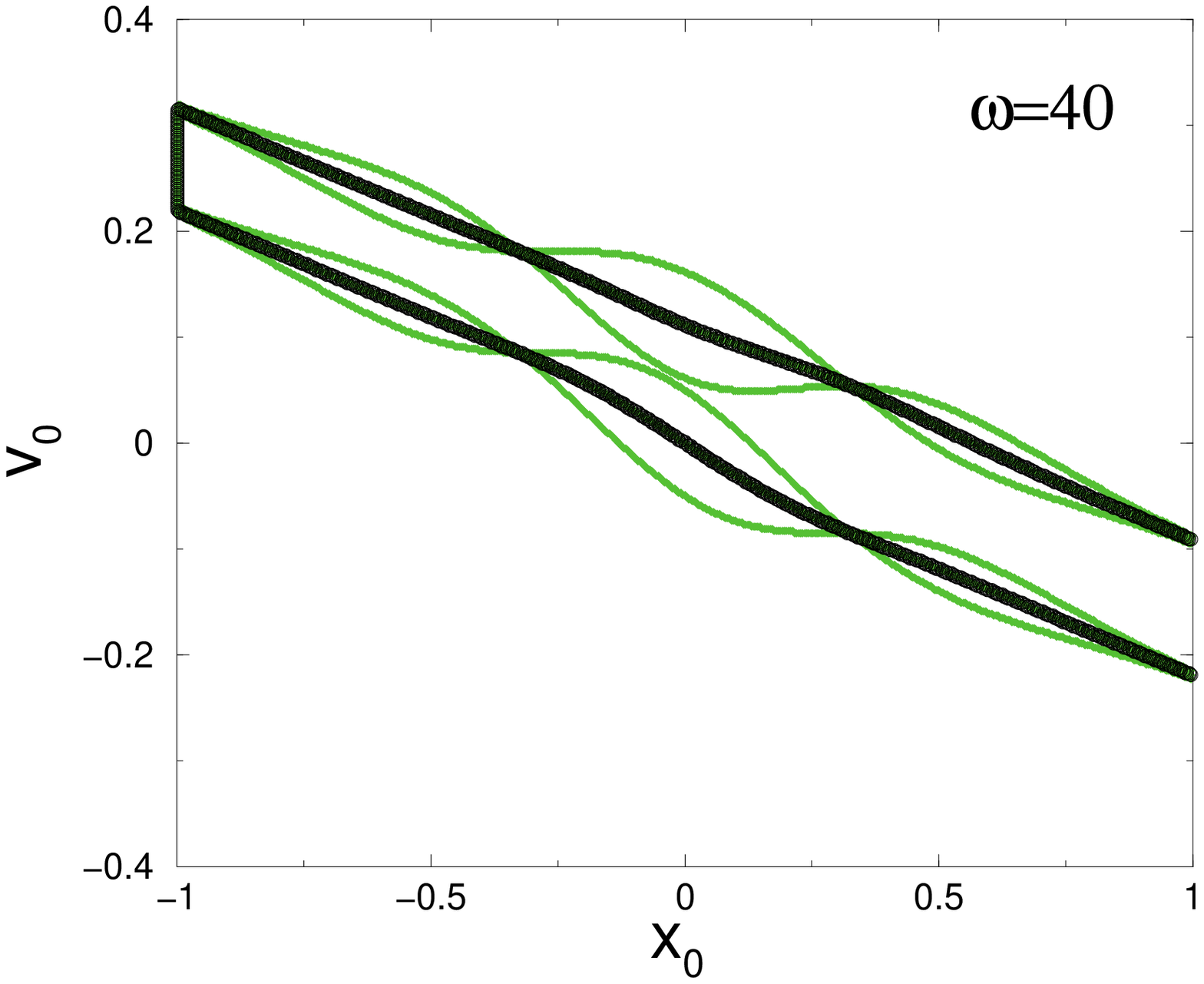}
 \end{minipage}%
 \caption{The boundary of the basin of attraction of $x \simeq 0.3266$ of the time-dependent system, for various phases,(thin lines) are compared to the one of the time-independent effective system (heavy line), for different frequencies. The values of the phase are $\phi=0$, $\pi/2$, $\pi$ and $3\pi/2$. \label{fig3}}
\end{figure}

It was demonstrated that some qualitative properties of the time-dependent 
system are described
by properties of an appropriate approximate time-independent system. 
It is of interest to see whether one can obtain also quantitative
results using the high-frequency perturbation theory. 
The boundaries of one of the basins of attraction of $x \simeq 0.3266$
of the time-dependent system are compared to the ones resulting from the
approximated time-independent system in Fig.~\ref{fig3}.
The motion of the time dependent system depends also on the phase 
of the oscillating force at $t=0$. It is important to note that this
comparison is fairly naive since the initial values $x_0$ and $v_0$
are only approximately equal to $X_0$ and $\dot{X}_0$ of the
time independent system. 
Fig.~\ref{fig3} shows that the boundary of the basin of attraction
of the time-dependent system fluctuates around the corresponding
boundary of the effective time-independent system when the phase
of the field is varied. This is not surprising since it is clear that 
the basins of attraction must depend on this phase.

As was mentioned earlier, the comparison in Fig.~\ref{fig3} is naive,
in the sense that the coordinates $x(0),v(0)$ are {\em not}
the same as the coordinates $X(0),\dot{X} (0)$ of the time-independent system.
However, the results presented in Fig.~\ref{fig3} still demonstrate
that at high frequencies, the basins of attraction of the time-dependent 
system can be approximated by those of time-independent effective 
ones. The size of the fluctuations seem
to decrease when the frequency is increased.

To obtain a better quantitative correspondence between the results
obtained using the high-frequency perturbation theory and the
numerical results for the time-dependent system, one has to account for
the difference between the slow coordinates $X,\dot{X}$ and $x,v$.
The connection between $x(t)$ and $X(t)$ (and also $\dot{X}(t)$)
is given by Eq. (\ref{ansatz}).
One can use the high frequency perturbation theory to obtain an
expansion for $\xi$ and then substitute $t=0$. This leads to
an equation for the initial value $x_0=x(0)$ in terms of 
$X_0=X(0), \dot{X}_0=\dot{X}(0)$.
Similarly, by differentiating with respect to time at $t=0$ one obtains $v_0=v(0)$. The calculation results in
\begin{eqnarray}
\label{initialtr}
x_0 & = & X_0 + \frac{A}{m \omega^2} e^{- \beta X_0^2} \cos (\phi) f(X_0)
+ O(\omega^{-3}) \nonumber \\
v_0 & = & \dot{X}_0 - \frac{A}{m \omega} e^{- \beta X_0^2} \sin (\phi) f(X_0)
\nonumber \\
& - & \frac{A}{m \omega^2} e^{- \beta X_0^2}\dot{X}_0 \cos (\phi) g(X_0) \nonumber \\
 & - &  \frac{\alpha A}{m^2 \omega^2} e^{- \beta X_0^2}\cos (\phi) f(X_0) + O(\omega^{-3})
\end{eqnarray}
where $f(X_0)= k \cos (k X_0) - 2 \beta X_0 \sin (k X_0)$,
while $g(X_0) = \left[ 4 \beta^2 X_0^2 - k^2 - 2 \beta\right] \sin(k X_0)- 4 \beta X_0 k \cos (k X_0)$.
From Eq. (\ref{initialtr}) it is clear that the fluctuations in
$v_0$ are of order $\omega^{-1}$ when the phase is varied,
while the fluctuations in $x_0$ scale only as $\omega^{-2}$.
This is in agreement with the fluctuations presented in Fig.~\ref{fig3}.

\begin{figure}[t]
 \begin{minipage}[c]{0.45\textwidth}
	\centering \includegraphics[width=8cm,height=5cm]{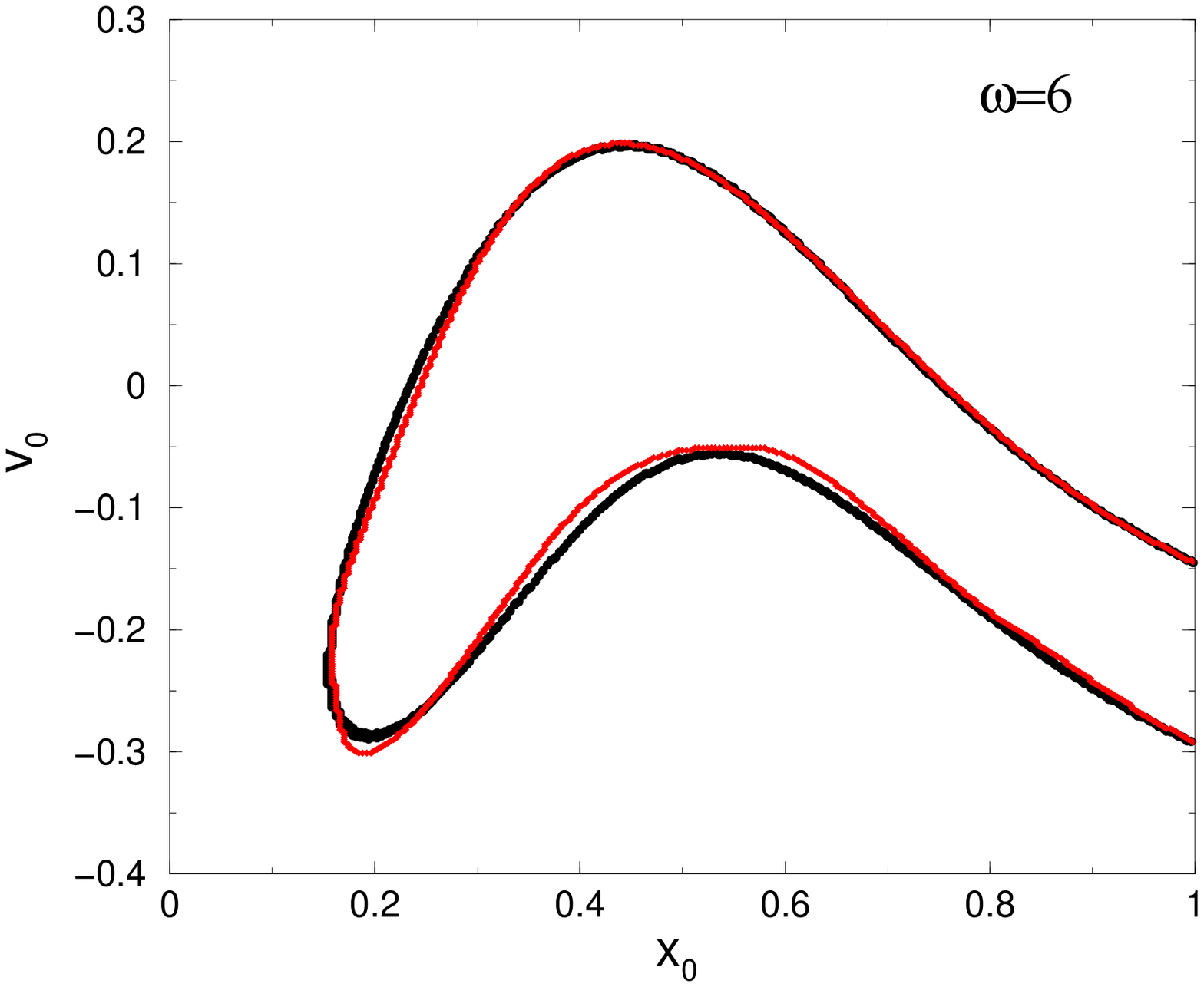}
 \end{minipage}\\[5pt]
 \begin{minipage}[c]{0.45\textwidth}
	\centering \includegraphics[width=8cm,height=5cm]{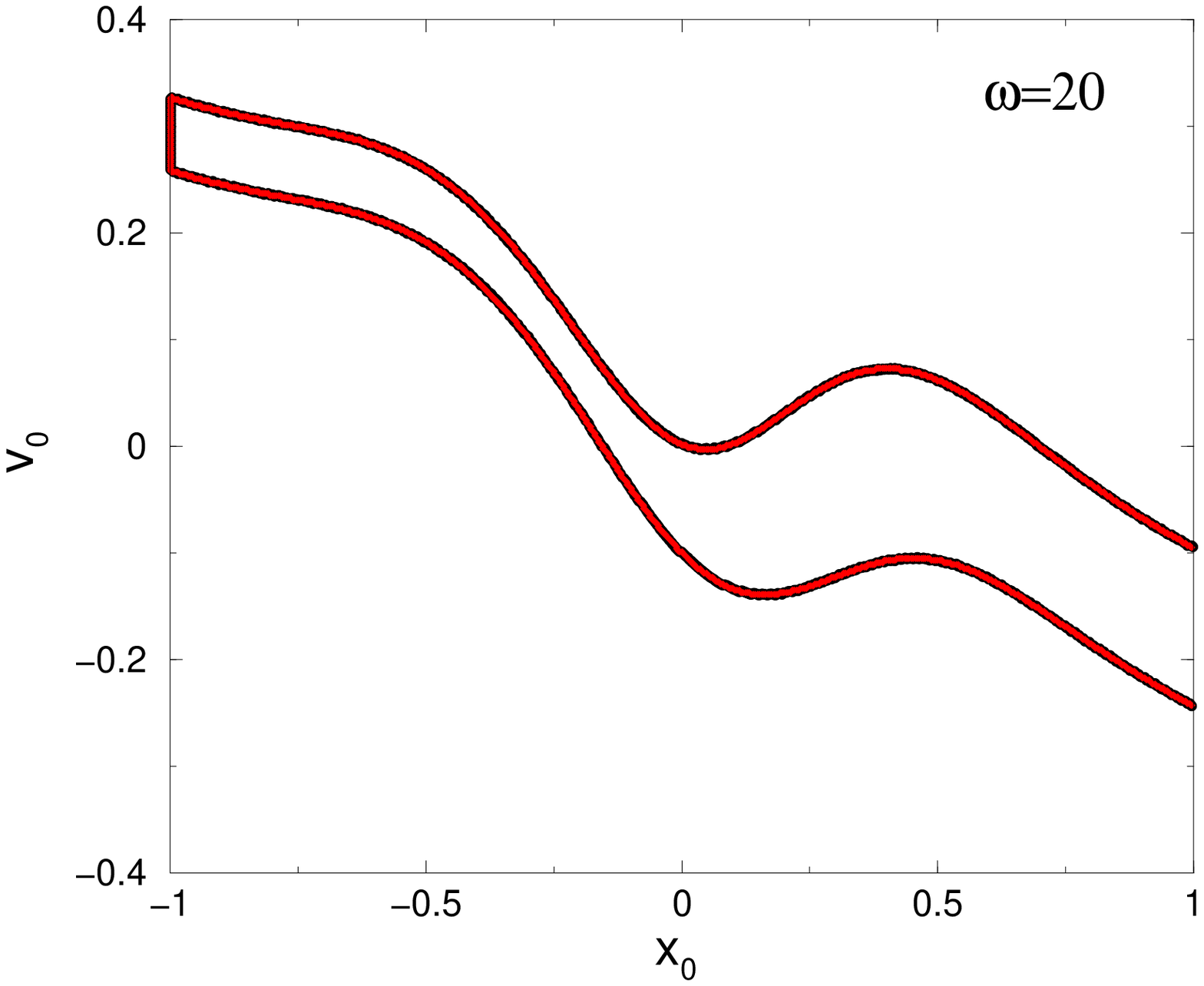}
 \end{minipage}%
 \caption{Comparison between the boundary of the basin of attraction
of $x\simeq 0.3266$ of the time-independent system (heavy line) and the boundary of the time-dependent one (thin line), for different frequencies, where $\phi=\pi/2$ while the rest of the parameters are those of Fig. \protect{\ref{fig2}}.\label{fig4}}
\end{figure}

To test the perturbation theory more quantitatively,
the boundary of the basin of attraction of the effective
time independent system was mapped back to the original coordinates
using Eq. (\ref{initialtr}) and compared to the numerical results
obtained for the time-dependent system. To avoid complicated graphs,
we only present the comparison for the phase $\phi=\pi/2$.
The results are presented in Fig.~\ref{fig4}.
It is clear that the agreement is excellent. Only a small difference
is seen for $\omega=6$ while for $\omega=20$ the difference between the 
boundaries cannot be observed on the plot.
This demonstrates that the high-frequency perturbation theory
can be used to obtain quantitative results and not only
qualitative ones.
 
In this article, we have used a high frequency perturbation theory
to describe the motion of a rapidly driven classical
particle in the presence of friction. In this perturbation theory
the motion is separated into a slow part and rapid oscillations 
around it (see Eqs. (\ref{ansatz})-(\ref{slow})).
An equation of motion for the slow coordinate, accurate to order $\omega^{-3}$,
was obtained. The oscillations of the particle around the slow solution 
were calculated as well.
The slow motion is found to be approximately described, to the order $\omega^{-2}$,
by the motion of a particle in an effective potential with
friction. This suggests that after a long time the particle will
be found at the minimum of this potential, a fact which is
confirmed numerically also for the time-dependent system given by Eq. (\ref{example}).
Due to the dissipation, these minima are surrounded by basins of
attraction, which include the initial phase space points that
flow to those minima after a long time.
The numerical results, presented in Figs. \ref{fig2} and \ref{fig3}
demonstrate that some of the qualitative features
of the basins of attraction can be understood by considering the
simpler time-independent effective system
with the potential of Fig.~\ref{fig1}. It was also
shown, in Fig.~\ref{fig4}, that, by a more careful analysis,
 one can obtain excellent quantitative agreement
between the motion generated by the time-dependent equation of
motion and the motion generated by the corresponding time-independent one.

The results presented in this article suggest that the high frequency
perturbation theory can be used to obtain both qualitative and
quantitative understanding of the dynamics of a classical rapidly driven
particle in the presence of friction.
The friction effectively dissipates the energy (only) of
the slow motion (up to the order $\omega^{-2}$).
\begin{acknowledgments}
This research was supported in part by the US-Israel Binational
Science
Foundation (BSF),
 by the Minerva Center of Nonlinear Physics of Complex Systems,
by the Fund for Promotion of Research at the Technion and by
Shlomo Kaplansky Academic Chair.
\end{acknowledgments}

\typeout{References}

\end{document}